\documentclass[osajnl,twocolumn,eqsecnum,showpacs]{revtex4}

\usepackage{amsfonts,amssymb,amsbsy,bm,graphicx}

\begin{document}


\title{A vectorlike representation of multilayers}

\author{Alberto G. Barriuso, Juan J. Monz\'on,
and Luis L. S\'anchez-Soto}

\affiliation{Departamento de \'Optica,
Facultad de F\'{\i}sica,
Universidad Complutense,
28040 Madrid, Spain}

\author{Jos\'e F. Cari\~{n}ena}
\affiliation{Departamento de F\'{\i}sica Te\'orica,
Facultad de Ciencias, Universidad de Zaragoza,
50009 Zaragoza, Spain}

\begin{abstract}
We resort to the concept of turns to provide a
geometrical representation of the action of any
lossless multilayer, which can be considered as
the analogous in the unit disk to the sliding
vectors in Euclidean geometry. This construction
clearly shows the peculiar effects arising in the
composition of multilayers. A simple optical
experiment revealing the appearance of the Wigner
angle is analyzed in this framework.
\end{abstract}


\maketitle

\newpage

\section{Introduction}

The search for mathematical entities that could
describe physical phenomena has always been a top
priority. For example, the need to describe a direction
in space, combined with the use of geometry to
approach physical problems, brought forth the concept
of a vector.

The idea that the complex numbers have a geometrical
interpretation as vectors lying in a plane, led to
Hamilton to introduce quaternions with the aim of
being useful for the analysis of three-dimensional
space~\cite{Ham53}. The price to be paid is that the
composition of quaternions is not commutative. Soon
after that, it became clear that rotations can be
advantageously represented by unit quaternions.

A notion closely related to Hamilton treatment is
that of turns~\cite{Bid81}. The turn associated with a
rotation of axis $\hat{\mathbf{n}}$ and angle $\vartheta$ is
a directed arc of length $\vartheta/2$ on the great circle
orthogonal to $\hat{\mathbf{n}}$ on the unit sphere. By
means of these objects, the composition of rotations is
described through a parallelogramlike law: if these
turns are translated on the great circles, until the
head of the arc of the first rotation coincides with
the tail of the arc of the second one, then the turn
between the free tail and the head is associated with
the resultant rotation. Hamilton turns are thus analogous
for spherical geometry to the sliding vectors in
Euclidean geometry.

In recent years many concepts of geometrical nature
have been introduced to gain further insights into
the behavior of layered media. The algebraic basis
for these developments is the fact that the transfer
matrix associated with a lossless multilayer is an
element of the group SU(1,~1), which is locally
isomorphic to the  Lorentz group SO(2,~1) in
(2+1) dimensions. This leads to a natural and complete
identification between reflection and transmission
coefficients and the parameters of the corresponding
Lorentz transformation~\cite{Mon99a,Mon99b}.

In an appealing paper, Ju\'arez and Santander~\cite{Jua82}
developed a generalization of Hamilton turns to
the Lorentz group, while Simon, Mukunda, and
Sudarshan~\cite{Sim89a,Sim89b} worked out an equivalent
algebraic approach for SU(1,~1), in which they
introduce a noncommutative geometrical addition
for these hyperbolic turns that reproduces the composition
law of the group; i. e.,  for both the reflection and
transmission coefficients. The latter coefficients
seem to be (almost) ignored in the literature.
The goal of this paper is precisely to show how this
formalism affords a very intuitive image of multilayer
optics.

Of course, since the amplitude and phase of a light
beam can be conveniently represented by a (Euclidean)
vector, the properties of layered media have been
represented by graphical constructions. These
visualization tools are commonly in use in thin-film
design and include the Smith chart, the admittance
diagram~\cite{Mac86}, the (reflectance) circle
diagrams~\cite{Apf72}, and the vector method~\cite{Hea91},
among others. However, the formalism of turns is
intrinsic and does not rely on any specific
representation for the light waves. In other words,
once the turn for a multilayer is known, his action
on any light state is fully determined.  We emphasize
that, although, these techniques can be used for
quantitative calculations, they cannot compete
nowadays with modern software design, and
their great value is in the visualization of the
characteristics of a multilayer.

This paper is organized as follows: In Section 2
we present some details of how the action of any
multilayer can be seen as a geometrical motion in
the unit disk. Every one of these motions can be decomposed
in terms of two reflections, which justifies the idea
of turn introduced in Section 3, where their composition
law is also introduced via a parallelogram law, in
close analogy with what happens for sliding vectors
in Euclidean geometry. The noncommutative character
of this law leads to interesting phenomena, such as
the appearance of extra phases in the composition of
multilayers, which is examined in Section 4, giving a
simple though nontrivial example that illustrates clearly
how this geometrical scenario works in practice.

\section{Multilayer action in the unit disk}

The theory of reflection and transmission of light by
stratified planar structures is of wide interest in optics.
A considerable amount of theoretical work has been done on
this topic, a detailed discussion of which can be found in
a number of books~\cite{Hea91,Mac86,Bre60,Lek87,Aza87,Yeh88}.
Nowadays, the standard method uses a matrix
representation of the field in each medium, as
pioneered by Abel\`es~\cite{Abe48}. In this paper, we
follow the elegant approach developed by Hayfield and
White~\cite{Hay64}.

We deal with a stratified structure that consists
of a stack of plane-parallel layers sandwiched
between two semi-infinite ambient ($a$) and
substrate ($s$) media that we shall assume
to be identical, in order to simplify as much
as possible the calculations. Hereafter all the
media are supposed to be lossless, linear,
homogeneous, and isotropic. We choose the $Z$ axis
perpendicular to the boundaries and directed
as in Fig.~1.

\begin{figure}
\centering
\resizebox{0.75\columnwidth}{!}{\includegraphics{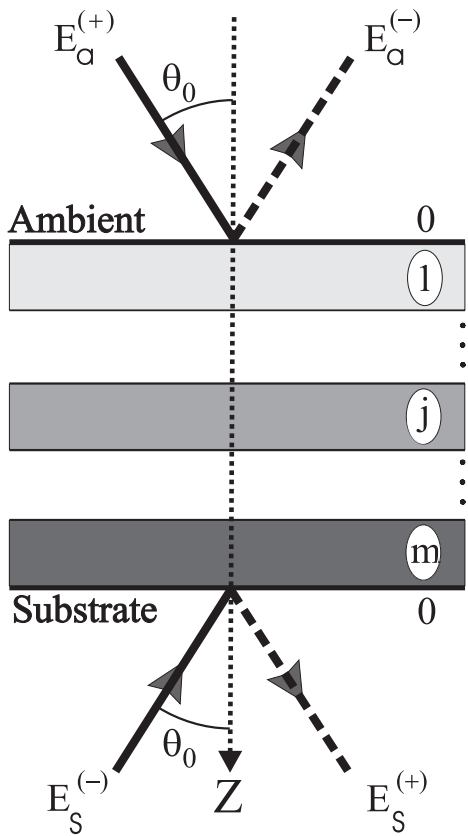}}
\caption{Wave vectors of the input $[E_{a}^{(+)}$ and
$E_{s}^{(-)}]$ and output $[E_{a}^{(-)}$ and $E_{s}^{(+)}]$ fields
in a multilayer sandwiched between two identical semi-infinite
ambient  and substrate media.}
\end{figure}

A monochromatic linearly polarized plane
wave falls from the ambient making an angle
$\theta_0$ with the normal to the first
interface and with an amplitude $E_{a}^{(+)}$.
The electric field is either in the plane of
incidence ($p$ polarization) or perpendicular to
the plane of incidence ($s$ polarization).
Since the multilayer has two input channels,
we consider as well another plane wave of the
same frequency and polarization, and with amplitude
$E_{s}^{(-)}$, incident from the substrate at
the same angle $\theta_0$. This guarantees that
all the multiple reflected and transmitted waves
superpose with the same wave vectors. The overall
output fields in the ambient and the substrate
will be denoted $E_{a}^{(-)}$ and $E_{s}^{(+)}$,
respectively. In the common experimental situation,
there is no light incident from the substrate
[$E_{s}^{(-)} = 0$], although there are relevant
situations in which both input fields are present
(e. g., at the beam splitter in a Michelson
interferometer).

The field amplitudes at each side of the
multilayer are related by the linear relation
\begin{equation}
\label{Evec}
\left (
\begin{array}{c}
E_a^{(+)} \\
E_a^{(-)}
\end{array}
\right )
= \bm{\mathsf{M}}_{as}
\left (
\begin{array}{c}
E_s^{(+)} \\
E_s^{(-)}
\end{array}
\right ) ,
\end{equation}
where the multilayer transfer matrix
$\bm{\mathsf{M}}_{as}$ can be shown
to be~\cite{Yeh88}
\begin{equation}
\label{Mlossless}
\bm{\mathsf{M}}_{as} = \left [
\begin{array}{cc}
1/T_{as} & R _{as}^\ast/T_{as}^\ast \\
R_{as}/T_{as} & 1/T_{as}^\ast
\end{array}
\right ] \equiv \left [
\begin{array}{cc}
\alpha & \beta \\
\beta^\ast & \alpha^\ast
\end{array}
\right ] .
\end{equation}
Here the complex numbers $R_{as}$ and $T_{as}$,
which can be expressed as
\begin{equation}
R_{as}  =  | R_{as} | \exp (i \rho) ,
\qquad
T_{as}  =  | T_{as} | \exp (i \tau) ,
\nonumber
\end{equation}
are, respectively, the overall reflection and
transmission coefficients for a wave incident
from the ambient. Because $ |R _{as}|^2 +
|T _{as}|^2 =1$, we have $\det \bm{\mathsf{M}}_{as}
= |\alpha |^2 - |\beta| ^2 =1$, and then
$\bm{\mathsf{M}}_{as}$ belongs to the group
SU(1,~1).

In Ref.~\cite{Yon02} we have proposed viewing
the multilayer action in a relativisticlike
framework, giving a formal equivalence between
the fields in Eq.~(\ref{Evec}) and space-time
coordinates in a $(2+1)$-dimensional space.
These coordinates verify $(x^{0})^{2} - (x^{1})^{2}
- (x^{2})^{2} = 1$, which defines a unit two-sheeted
hyperboloid characteristic of the group SO(2,~1).
If one uses stereographic projection taking
the south pole as projection center, the upper
sheet of the unit hyperboloid is projected into
the unit disk, the lower sheet into the external
region, while the infinity goes to the boundary
of the unit disk.

The geodesics in the hyperboloid are intersections
with the hyperboloid of planes passing through the
origin. Consequently, hyperbolic lines are obtained
from these by stereographic projection and they
correspond to circle arcs that orthogonally cut
the boundary of the unit disk.

In many instances (e.g., in polarization optics~\cite{Aza87})
we are interested in the transformation properties of field
quotients rather than the fields themselves. Therefore,
it seems natural to consider the complex numbers
\begin{eqnarray}
\label{defz}
z_s   =  \frac {E_s^{(-)}}{E_s^{(+)}}\ ,
\qquad
z_a  =   \frac {E_a^{(-)}}{E_a^{(+)}}\ .
\end{eqnarray}
The action of the multilayer given in Eq.~(\ref{Mlossless})
can be then seen as a function  $z_a = f(z_s)$ that can
be appropriately called the multilayer transfer
function~\cite{Ohl00}.

From a geometrical viewpoint, this function defines a
transformation of  the complex plane, mapping the
point $z_s$ into the point $z_a$, according
to~\cite{Mon02}
\begin{equation}
\label{accion}
z_a = \Phi [\mathsf{M}_{as} , z _{s}] =
\frac{\beta^\ast +\alpha^\ast z_s}
{\alpha + \beta z_s} \ ,
\end{equation}
which is a bilinear or M\"{o}bius transformation.
One can check that the unit disk, the external
region and the unit circle remain invariant under
the multilayer action. Note also that, when
no light impinges from the substrate $z_s=0$
and then $z_a = R_{as}$. It is worth mentioning
that this approach is quite general, since
it provides the transformed of any point in
the unit disk for every values of the input
fields $E_{a}^{(+)}$ and $E_{s}^{(-)}$.

To classify the multilayer action it proves
convenient to work out the fixed points of
the mapping; that is, the field configurations
such that $z_a = z_s \equiv z_f$ in
Eq.~(\ref{accion})~\cite{San01}:
\begin{equation}
z_f = \Phi [\bm{\mathsf{M}}_{as} , z_f] ,
\end{equation}
whose solutions are
\begin{equation}
z_f = \frac{1}{2 \beta}
\left \{  -2 i \ \mathrm{Im}(\alpha) \pm
\sqrt{[\mathrm{Tr} ( \bm{\mathsf{M}}_{as} )]^2 -4}
\right \} .
\end{equation}
When $ [\mathrm{Tr} ( \bm{\mathsf{M}}_{as} )] ^2
< 4$ the multilayer action is elliptic and
it has only one fixed point inside the unit
disk. Since in the Euclidean geometry a rotation
is characterized for having only one invariant
point, this multilayer action can be
appropriately called a hyperbolic rotation.

When $ [ \mathrm{Tr} ( \bm{\mathsf{M}}_{as} )]^2 > 4$
the action is hyperbolic and it has two fixed
points, both on the boundary of the unit disk.
The geodesic line joining these two fixed points
remains invariant and thus, by analogy with the
Euclidean case, this action will be called a
hyperbolic translation.

Finally, when $ [ \mathrm{Tr} (\bm{\mathsf{M}}_{as}) ]^2
= 4$ the multilayer action is  parabolic and it has only
one (double) fixed point on the boundary of the unit disk.

Here we will be concerned only with the case
$ [ \mathrm{Tr} ( \bm{\mathsf{M}}_{as} )]^2 > 4$, since
it is known that any element of SU(1,~1) can be
written (in many ways) as the product of two hyperbolic
translations~\cite{Sim89b}.  The axis of the hyperbolic
translation is the geodesic line joining the two fixed
points. A point on the axis will be translated to another
point, a (hyperbolic) distance~\cite{Bea83}
\begin{equation}
\zeta = 2 \ln   \left ( \frac{1}{2} \left \{
\mathrm{Tr} (\bm{\mathsf{M}}_{as}) + \sqrt{[
\mathrm{Tr} (\bm{\mathsf{M}}_{as}) ]^2 - 4}
\right \} \right )
\end{equation}
along the axis.

\section{Hyperbolic turns and their composition}

In Euclidean geometry, a translation of magnitude $\zeta$
along a line $\gamma$ can be seen as the product of
two reflections in any two straight lines orthogonal to
$\gamma$, separated a distance $\zeta/2$. This idea can be
translated much in the same way to the unit disk, once
the concepts of line and distance are understood in the
hyperbolic sense. In consequence, any pair of points $z_1$
and $z_2$ on the axis of the translation $\gamma$ at a
distance $\zeta/2$  can be chosen as intersections of
$\Gamma_1$ and $\Gamma_2$ (orthogonal lines to $\gamma$)
with $\gamma$. It is then natural to associate to the
translation an oriented segment of length $\zeta/2$ on
$\gamma$, but otherwise free to slide on $\gamma$
(see Fig.~2). This is analogous to Hamilton's turns, and
will be called a hyperbolic turn $\mathbb{T}_{\gamma, \zeta/2}$.

\begin{figure}
\centering
\resizebox{0.75\columnwidth}{!}{\includegraphics{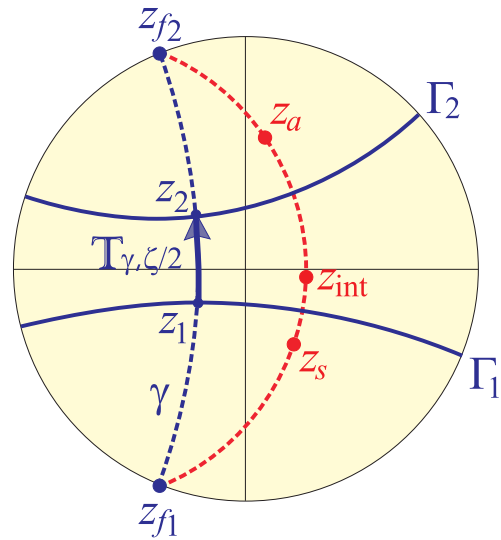}}
\caption{Representation of the sliding turn $\mathbb{T}_{\gamma,
\zeta/2}$ in terms of two reflections in two lines $\Gamma_1$ and
$\Gamma_2$ orthogonal to the axis of the translation $\gamma$,
which has two fixed points $z_{f1}$ and $z_{f2}$. The
transformation of a typical off axis point $z_s$ is also shown.}
\end{figure}

Note that using this construction, an off-axis point
such as $z_s$ will be mapped by these two reflections
(through an intermediate point $z_{\mathrm{int}}$) to
another point $z_a$ along a curve equidistant to the
axis. These other curves,  unlike the axis of translation,
are not hyperbolic lines. The essential point is
that once the turn is known, the transformation of
every point in the unit disk is automatically established.

Alternatively, we can formulate the concept of turn
as follows. Let $\bm{\mathsf{M}}_{as}$ be a hyperbolic
translation with $ \mathrm{Tr}  (\bm{\mathsf{M}}_{as} )  $
positive (equivalently, $\mathrm{Re} (\alpha ) > 1$). Then,
$\bm{\mathsf{M}}_{as}$ is positive definite and one
can ensure that its square root exists and reads as
\begin{equation}
\sqrt{\bm{\mathsf{M}}_{as}} =
\frac{1}{\sqrt{2 [ \mathrm{Re} (\alpha )
+ 1 ]}}
\left [
\begin{array}{cc}
\alpha + 1 &  \beta \\
\beta^\ast & \alpha^\ast + 1
\end{array}
\right ] .
\end{equation}
This matrix has the same fixed points as $\bm{\mathsf{M}}_{as}$,
but the translated distance is just half the induced by
$\bm{\mathsf{M}}_{as}$; i.e.,
\begin{equation}
\zeta ( \bm{\mathsf{M}}_{as}) =
2 \zeta (\sqrt{\bm{\mathsf{M}}_{as}}) .
\end{equation}
This suggests that the matrix $ \sqrt{\bm{\mathsf{M}}_{as}}$
can be appropriately associated to the turn
$\mathbb{T}_{\gamma, \zeta/2}$ that represents the
translation induced by $\bm{\mathsf{M}}_{as}$. Therefore, we
symbolically write
\begin{equation}
\label{mapsto}
\mathbb{T}_{\gamma, \zeta/2}\mapsto
\sqrt{\bm{\mathsf{M}}_{as}} .
\end{equation}

One may be tempted to extend the Euclidean
composition of concurrent vectors to the problem
of hyperbolic turns. Indeed, this can be done quite
straightforwardly~\cite{Jua82}. Let us consider
the case of the composition of two of these multilayers
represented by the matrices $\bm{\mathsf{M}}_1$ and
$\bm{\mathsf{M}}_2$ (for simplicity, we shall henceforth
omit the subscript $as$) of parameters $\zeta_1$
and $\zeta_2$ along intersecting axes $\gamma_1$
and $\gamma_2$, respectively. Take the associated
turns $\mathbb{T}_{\gamma_1, \zeta_1/2}$ and
$\mathbb{T}_{\gamma_2, \zeta_2 /2}$ and slide
them along $\gamma_1$ and $\gamma_2$ until they
are ``head to tail". Afterwards, the turn determined
by the free tail and head is the turn associated to
the resultant, which represents thus a translation of
parameter $\zeta$ along the line $\gamma$. This construction
is shown in Fig.~3, where the noncommutative
character is also evident.

\begin{figure}
\centering
\resizebox{0.75\columnwidth}{!}{\includegraphics{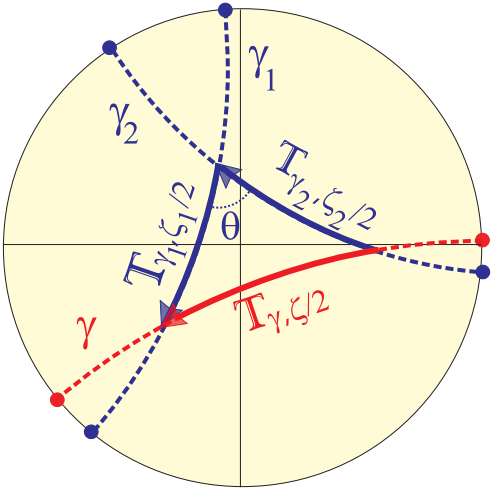}}
\caption{Composition of two hyperbolic turns
$\mathbb{T}_{\gamma_1, \zeta_1/2}$ and $\mathbb{T}_{\gamma_2,
\zeta_2/2}$ by using a parallelogramlike law when the axes
$\gamma_1$ and $\gamma_2$ of the translations intersect.}
\end{figure}

In Euclidean geometry, the resultant of this parallelogram law
can be quantitatively determined by a direct application of the
cosine theorem. For any hyperbolic triangle with sides of lengths
$\zeta_1$ and $\zeta_2$ that make an angle $\theta$, we take
the expression from any standard book on hyperbolic
geometry~\cite{Bea83}
\begin{equation}
\label{hcos}
\cosh \zeta = \cosh \zeta_1
\cosh \zeta_2 + \sinh \zeta_1
\sinh \zeta_2  \cos \theta ,
\end{equation}
where $\theta$ is the angle between both sides. Moreover,
for future use we quote that the (hyperbolic) area $\Omega$
of the geodesic triangle is
\begin{equation}
\label{harea}
\tan (\Omega/2) = \frac{\tanh ( \zeta_1/2) \tanh ( \zeta_2/2)
\sin \theta}{1 - \tanh ( \zeta_1/2) \tanh ( \zeta_2/2)
\cos \theta} .
\end{equation}

\section{Application: revealing the Wigner
angle in the unit disk}

To show how this formalism can account for
the existence of peculiar effects in the
composition of multilayers, we address here
the question of the Wigner angle in the unit
disk and propose a simple optical experiment
to determine this angle.

The Wigner angle emerges in the study of the
composition of two noncollinear pure boosts in
special relativity: the combination of two such
successive boosts cannot result in a pure boost,
but  renders an additional pure rotation, usually
known as the Wigner rotation~\cite{BE85}  (sometimes
the name of Thomas rotation~\cite{Jac75,Ung89}
is also used). In other words, boosts are not
a group.

To fix the physical background, consider three frames of reference
$K$, $K^\prime$ and $K^{\prime \prime}$. Frames $K$-$K^\prime$
and $K^\prime$-$K^{\prime \prime}$ have parallel respective axes.
Frame $K^{\prime \prime}$  moves with uniform velocity
$\mathbf{v}_2$ with respect to  $K^\prime$, which in turn moves
with velocity $ \mathbf{v}_1$ relative to $K$. The Lorentz
transformation that connects $K$ with $K^{\prime \prime}$ is given
by the product $\bm{\mathsf{L}}_1 (\mathbf{v}_1) \bm{\mathsf{L}}_2
(\mathbf{v}_2) $, which can be decomposed as
\begin{equation}
\label{bcom}
\bm{\mathsf{L}}_1(\mathbf{v}_1)
\bm{\mathsf{L}}_2(\mathbf{v}_2)
= \bm{\mathsf{L}}_{(12)} (\mathbf{v})
\bm{\mathsf{R}}(\Psi) ,
\end{equation}
where one must be careful in operating just in
the same order as written in Eq.~(\ref{bcom}).
In words, this means that an observer in $K$
sees the axes of  $K^{\prime \prime}$ rotated
relative to  the observer's own axes by a
Wigner rotation described by $\bm{\mathsf{R}} (\Psi)$.
The explicit expression for the axis and angle
of this rotation can be found, e.g. in Ref.~\cite{BE85}
and will be worked out below from the perspective
of the multilayer action in the unit disk.

\begin{figure}
\centering
\resizebox{0.75\columnwidth}{!}{\includegraphics{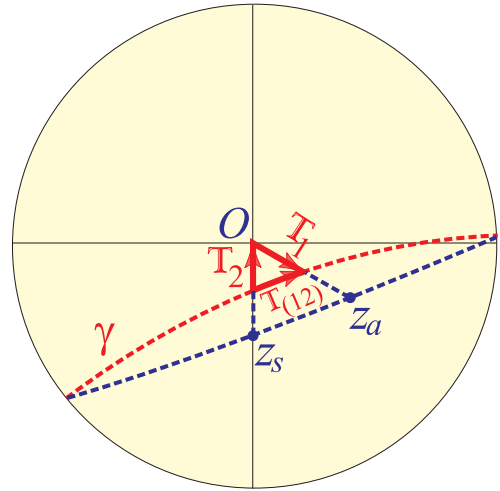}}
\caption{Composition of two multilayers represented by Hermitian
matrices $\bm{\mathsf{H}}_1$ and $\bm{\mathsf{H}}_2$.
$\bm{\mathsf{H}}_2$ maps the point $z_s= - R_2$ into the origin,
while $\bm{\mathsf{H}}_1$ maps the origin into $z_a=R_1$. We show
also the associated turns $\mathbb{T}_1$ and $\mathbb{T}_2$, as
well as the resulting one $\mathbb{T}_{(12)}$ obtained via the
parallelogram law. The composite multilayer $\bm{\mathsf{H}}_1
\bm{\mathsf{H}}_2$ transforms the point $z_s$ into $z_a$. The data
of the corresponding multilayers are given in the text.}
\end{figure}

First, we observe that any matrix $\bm{\mathsf{M}} \in$ SU(1,~1)
can be expressed in a unique way in the form
\begin{equation}
\bm{\mathsf{M}} = \bm{\mathsf{H}}
\bm{\mathsf{U}} ,
\end{equation}
where $\bm{\mathsf{H}}$ is positive definite
Hermitian and $\bm{\mathsf{U}}$ is unitary.
One can check by simple inspection that the
explicit form of this (polar) decomposition reads
as~\cite{Mon99c}
\begin{eqnarray}
\label{polar}
\bm{\mathsf{M}} & =  & \bm{\mathsf{H}}
\bm{\mathsf{U}}
=
\left [
\begin{array}{cc}
\ 1/|T| \  & \ R^\ast/|T| \ \\
\ R/|T| \ & \ 1/|T| \
\end{array}
\right ] \nonumber \\
& \times &
\left [
\begin{array}{cc}
\exp (- i \tau ) & 0 \\
0 & \exp (i \tau )
\end{array}
\right ] .
\end{eqnarray}
The component $\bm{\mathsf{H}}$ is equivalent to
a pure boost, while $\bm{\mathsf{U}}$ is equivalent
to a spatial rotation.

It is clear from Eq.~(\ref{polar}) that
$[ \mathrm{Tr}  (\bm{\mathsf{H}} )]^2 > 4$,
so it represents a hyperbolic translation.
Moreover, one can check that its associated
fixed points are diametrically located on
the unit circle and so, the axis of this
translation is precisely the diameter joining
them. By writing now
\begin{equation}
R  =   \tanh (\zeta/2) \exp(i  \rho ) ,
\qquad
T  =   \mathrm{sech} (\zeta/2)
\exp( i \tau) ,
\end{equation}
one can easily check that the matrix $\bm{\mathsf{H}} $
in Eq.~(\ref{polar}) transforms the origin into the
complex point $R$, that is,
\begin{equation}
\label{clever}
\Phi [\bm{\mathsf{H}}, 0 ] = R ,
\qquad
\Phi [\bm{\mathsf{H}}^{-1}, R ] =  0 .
\end{equation}

In complete analogy with Eq.~(\ref{bcom})
we compose now two multilayers represented by
Hermitian matrices $\bm{\mathsf{H}}_1$ and
$\bm{\mathsf{H}}_2$ (that is, with zero transmission
phase lag $\tau_1 = \tau_2 = 0$) and we get, after
simple calculations~\cite{Mon01a}
\begin{eqnarray}
\label{Htot}
\bm{\mathsf{H}}_1 \bm{\mathsf{H}}_2
& = &
\bm{\mathsf{H}}_{(12)} \bm{\mathsf{U}} =
\left [
\begin{array}{cc}
1/|T_{(12)}| & R_{(12)}^\ast/|T_{(12)}| \\
R_{(12)} /|T_{(12)}|&  1/|T_{(12)}|
\end{array}
\right ] \nonumber \\
& \times &
\left [
\begin{array}{cc}
\exp (- i \Psi /2 ) & 0 \\
0 & \exp ( i \Psi /2  )
\end{array}
\right ] ,
\end{eqnarray}
where
\begin{eqnarray}
\label{compo12}
&  \displaystyle
R_{(12)}    =
\frac{R _1 + R_2}{1+ R _1^\ast R_2} ,
\qquad
T_{(12)}   =
\frac{| T_1 T_2 |}{1+ R _1^\ast R_2 } , &
\nonumber \\
&  & \\
& \displaystyle
\frac{\Psi}{2} =  \arg [ T_{(12)}] =
\arg(1+ R_1 R _2^\ast) , &
\nonumber
\end{eqnarray}
and the subscripts 1 and 2 refer to the
corresponding multilayer. The appearance of
an extra unitary matrix in Eq.~(\ref{Htot}) is
the signature of a Wigner rotation in the
multilayer composition and, accordingly,
the Wigner angle $\Psi$ is just twice
the phase of the transmission coefficient
of the compound multilayer.

To view this Wigner angle in the unit disk, let
$z_s$ be the point in the substrate that is
transformed by the multilayer $\bm{\mathsf{H}}_2$
into the origin, and let $z_a$ be the result of
transforming the origin by $\bm{\mathsf{H}}_1$.
According to Eq.~(\ref{clever}), one has
\begin{equation}
\Phi [\bm{\mathsf{H}}_2, - R_2 ] = 0 ,
\qquad
\Phi [\bm{\mathsf{H}}_1, 0 ] = R_1 .
\end{equation}

Consider now the (geodesic) triangle defined by the points $z_s$,
$O$ and $z_a$ in Fig.~4. The general formula (\ref{harea}) gives
for this triangle
\begin{equation}
\Omega =  \Psi ,
\end{equation}
which confirms the geometric nature of this
Wigner angle, since it can be understood in terms
of the area (or equivalently, the anholonomy) of
a closed circuit~\cite{geomphas,Ara97}.

According to the ideas developed in Section 3,
we can reduce the multilayers $\bm{\mathsf{H}}_1$
and $\bm{\mathsf{H}}_2$ to the associated turns,
represented by arrows in Fig.~4. The ``head to tail"
rule applied to $\mathbb{T}_1$ and $\mathbb{T}_2$
(for simplicity, we omit in the subscripts of these
turns the corresponding parameters) immediately
gives the resulting turn $\mathbb{T}_{(12)}$.
However, note that, if we follow the formal prescription
shown in Eq.~(\ref{mapsto}) and ascribe $\mathbb{T}_1
\mapsto \sqrt{\bm{\mathsf{H}}_1}$ and $\mathbb{T}_2
\mapsto \sqrt{\bm{\mathsf{H}}_2}$, we conclude that
the composition law imposes
\begin{equation}
\mathbb{T}_{(12)} \mapsto
\sqrt{\bm{\mathsf{H}}_1 \bm{\mathsf{H}}_2} .
\end{equation}

\begin{figure}
\centering
\resizebox{0.75\columnwidth}{!}{\includegraphics{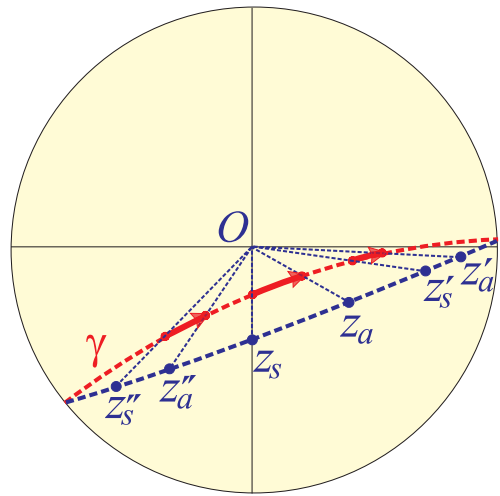}}
\caption{The same as in Fig.~4 but now the resulting turn
$\mathbb{T}_{(12)}$ has been slided to three different positions
along the axis. The corresponding points are also transformed by
$\bm{\mathsf{H}}_1 \bm{\mathsf{H}}_2 $. All the geodesic triangles
plotted  have the same hyperbolic area $\Psi$.}
\end{figure}

All these results are independent on the position of
the turn. In fact, in Fig.~5 we have put the turn
$\mathbb{T}_{(12)}$ in different positions along the
axis $\gamma$. In every position, we have drawn two
radii passing through the head and the tail of
$\mathbb{T}_{(12)}$ and taken on them twice the
hyperbolic distance from the origin. The pairs of
points obtained in this way (such as $z_s^\prime$ and
$z_a^\prime$) are transformed precisely by
$\bm{\mathsf{H}}_1 \bm{\mathsf{H}}_2$.
In other words, $\bm{\mathsf{H}}_1 \bm{\mathsf{H}}_2$ can
be decomposed in many ways as the composition of two Hermitian
matrices and every geodesic triangle $z_s^\prime \ O \ z_a^\prime$
has the same hyperbolic area $\Psi$.

It seems pertinent to conclude by showing an experimental
implementation of the data shown in Fig.~4.
To this end, we first recall~\cite{Aza87} that for a
single plate of refractive index $n_j$ and thickness $d_j$
imbedded in air ($n_0=1$) illuminated with a monochromatic
light of wavelength in vacuo $\lambda$, a standard calculation
gives the transfer matrix $\bm{\mathsf{M}}_{0j0}$ with
the following reflection and transmission
coefficients:
\begin{eqnarray}
\label{RT}
R_{0j0} & = &
\frac{r_{0j} [1 - \exp(- i 2 \delta_j) ]}
{1- r_{0j}^2  \exp(- i 2 \delta_j) } , \nonumber \\
& & \\
T_{0j0} & = &
\frac{t_{0j} t_{j0} \exp(- i \delta_j)}
{1- r_{0j}^2  \exp(- i 2 \delta_j) } ,
\nonumber
\end{eqnarray}
where $r_{0j}$ and $t_{0j}$ are the Fresnel reflection
and transmission coefficients at the interface $0j$ (which
applies to both $p$ and $s$ polarizations by the simple
attachment of a subscript $p$ or $s$) and $\delta_j$ is
the plate phase thickness
\begin{equation}
\delta_j = \frac{2 \pi}{\lambda} n_j d_j \cos \theta_j ,
\end{equation}
$\theta_j$ being the angle of refraction in the layer.
The transfer matrix for the coherent addition of $m$ of these
layers is
\begin{equation}
\bm{\mathsf{M}} = \prod_{j=1}^m \bm{\mathsf{M}}_{0j0} .
\end{equation}

\begin{figure}
\centering
\resizebox{0.75\columnwidth}{!}{\includegraphics{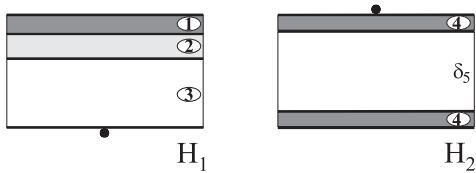}}
\caption{Scheme of two Hermitian multilayers $\bm{\mathsf{H}}_1$
and $\bm{\mathsf{H}}_2$. The compound multilayer
$\bm{\mathsf{H}}_1 \bm{\mathsf{H}}_2$ obtained putting together
these two components induces a Wigner rotation of angle $2 \tau$.}
\end{figure}

As shown in Fig.~6, we take as the first multilayer
$\bm{\mathsf{H}}_1$ the lossless system formed
by two thin films, one of zinc sulphide  (with refractive
index $n_1 = 2.3$ and thickness $d_1 = 80$~nm) and the other of
cryolite (with refractive index $n_2 = 1.35$ and
thickness $d_2 = 104$~nm), deposited on a glass substrate
(with refractive index $n_3 = 1.5$ and thickness $d_3 = 1.3$~mm),
and imbedded  in air. The light has a wavelength
in vacuo of $\lambda_0 =546$~nm and falls from
the ambient at normal incidence. Such a simple system
could be manufactured with standard evaporation techniques.

We have performed a computer simulation of the
performance of this multilayer $\bm{\mathsf{H}}_1$ using
a standard package, obtaining  $T_1 = 0.9055$
and  $R_1 = 0.3736 -  0.2014 i$, which in turn gives
$\tau_1 =0$ and $\rho_1= -0.4944$ \ rad.

Our second multilayer  $\bm{\mathsf{H}}_2$ is a symmetric
system formed by two films of zinc sulphide and thickness
$d_4 =  40$~nm separated by a spacer of air with a phase thickness
$\delta_5 = 3.707$~rad. For this subsystem we have $T_2 = 0.9399$
and $R_2 = 0.3413 i$, and therefore $\tau_2 =0$ and
$\rho_2= \pi/2$ \ rad.

When these two multilayer are put together by the marked
points in Fig.~6, the resulting one has a transmission
phase lag of $\tau = - 0.1361 \ $ rad, which is just
half the area of the geodesic triangle $z_s O  z_a$
in Fig.~4, as predicted by the theory.

In summary, we expect that the geometrical approach
presented here will be an interesting tool for representing
in a graphical way the multilayer action. Moreover,
the composition law of these turns allows for a clear
understanding of the nontrivial effects appearing
in the composition of multilayers.

We stress that the benefit of this approach lies not in
any inherent advantage in terms of efficiency in solving
problems in layered structures. Rather, we expect that
turns could provide a general and unifying tool to analyze
multilayer performance in an elegant  way that, in
addition, is closely related to other fields of physics.

\begin{acknowledgments}
We wish to thank Jos\'e Mar\'{\i}a  Montesinos and
Mariano Santander for enlightening discussions.

Corresponding author Luis L. S\'anchez-Soto
e-mail address is lsanchez@fis.ucm.es.
\end{acknowledgments}


\begin{thebibliography}{99}
\bibitem{Ham53}
W. R. Hamilton, \textit{Lectures on Quaternions}
(Hodges and Smith, Dublin, 1853).

\bibitem{Bid81}
L. C. Biedenharn and J. D. Louck,
\textit{Angular Momentum in Quantum Physics}
(Addison, Reading, MA 1981).

\bibitem{Mon99a}
J. J. Monz\'{o}n and L. L. S\'{a}nchez-Soto,
``Lossless multilayers and Lorentz transformations:
more than an analogy,''
Opt. Commun. \textbf{162}, 1-6 (1999).

\bibitem{Mon99b}
J. J. Monz\'{o}n and L. L. S\'{a}nchez-Soto,
``Fully relativisticlike formulation of multilayer optics,''
J. Opt. Soc. Am. A \textbf{16}, 2013-2018 (1999).

\bibitem{Jua82}
M. Ju\'arez and M. Santander,
``Turns for the Lorentz group,"
J. Phys. A  \textbf{15}, 3411-3424 (1982).

\bibitem{Sim89a}
R. Simon, N. Mukunda, and E. C. G. Sudarshan,
``Hamilton's Theory of Turns Generalized to Sp(2,R),"
Phys. Rev. Lett. \textbf{62}, 1331-1334 (1989).

\bibitem{Sim89b}
R. Simon, N. Mukunda, and E. C. G. Sudarshan,
``The theory of screws: A new geometric representation
for the group SU(1, 1),"
J. Math. Phys. \textbf{30}, 1000-1006 (1989).

\bibitem{Mac86}
H. A. Macleod, \textit{Thin-film Optical Filters}
(Adam Hilger, Bristol, UK, 1986).

\bibitem{Apf72}
J. H. Apfel, ``Graphics in Optical Coating Design,"
Appl. Opt. \textbf{11}, 1303-1312 (1972)

\bibitem{Hea91}
O. S. Heavens, \textit{Optical Properties of Thin Solid Films}
(Dover, New York, 1991).

\bibitem{Bre60}
L. M. Brekovskikh,\textit{ Waves in Layered Media}
(Academic, New York, 1960).

\bibitem{Lek87}
J. Lekner, \textit{Theory of Reflection}
(Dordrecht, The Netherlands, 1987).

\bibitem{Aza87}
R. M. A. Azzam and N. M. Bashara,
\textit{Ellipsometry and Polarized Light}
(North-Holland, Amsterdam, 1987).

\bibitem{Yeh88}
P. Yeh, \textit{Optical Waves in Layered Media}
(Wiley, New York, 1988)

\bibitem{Abe48}
F. Abel\`es, ``Sur la propagation des ondes
electromagnetiques dans les milieux statifi\'es,"
Ann. Phys. (Paris) \textbf{3}, 504-520 (1948)

\bibitem{Hay64}
P. C. S. Hayfield and G. W. T. White,
``An assessment of the stability of the Drude-Tronstad
polarized light method for the study of film growth on
polycrystalline metals," in
\textit{Ellipsometry in the Measurements of Surfaces and Thin Films},
E. Passaglia, R. R. Stromberg, and J. Kruger, eds.,
Natl. Bur. Stand. Misc. Publ. 256 (U.S.  GPO, Washington, D.C., 1964), pp. 157-200.

For a more recent review of the model
see Ref.~\cite{Aza87}, Sec. 4.6.

\bibitem{Yon02}
T. Yonte, J. J. Monz\'{o}n, L. L. S\'{a}nchez-Soto,
J. F. Cari\~{n}ena, and C. L\'opez-Lacasta,
``Understanding multilayers from a geometrical viewpoint,''
J. Opt. Soc. Am. A \textbf{19}, 603-609 (2002).

\bibitem{Ohl00}
I. Ohl\'{\i}dal and D. Franta,
\textit{Ellipsometry of Thin Film Systems},
Progress in Optics (Edited by E. Wolf)
\textbf{41}, 181-282 (North-Holland, Amsterdam, 2000).

\bibitem{Mon02}
J. J. Monz\'{o}n, T. Yonte, L. L. S\'{a}nchez-Soto, and J. F.
Cari\~{n}ena, ``Geometrical setting for the classification of
multilayers,'' J. Opt. Soc. Am. A \textbf{19}, 985-991 (2002).

\bibitem{San01}
L. L. S\'anchez-Soto, J. J. Monz\'on,
T. Yonte, and J. F. Cari\~{n}ena,
``Simple trace criterion for classification of multilayers,"
Opt. Lett. \textbf{26}, 1400-1402 (2001).

\bibitem{Bea83}
A. F. Beardon,
\textit{The Geometry of Discrete Groups}
(Springer, New York, 1983) Chap. 7.

\bibitem{BE85}
A. Ben-Menahem, ``Wigner's rotation revisited,"
Am. J. Phys. \textbf{53}, 62-66 (1985).

\bibitem{Jac75}
D. A. Jackson, \textit{Classical Electrodynamics}
(Wiley, New York, 1975).

\bibitem{Ung89}
A. A. Ungar, ``The relativistic velocity composition paradox
and the Thomas rotation,"
Found. Phys. \textbf{19}, 1385-1396 (1989).

\bibitem{Mon99c}
J. J. Monz\'{o}n and L. L. S\'{a}nchez-Soto,
``Origin of the Thomas rotation that arises
in lossless multilayers,''
J. Opt. Soc. Am. A  \textbf{16}, 2786-2792 (1999).

\bibitem{Mon01a}
J. J. Monz\'{o}n and  L. L. S\'{a}nchez-Soto,
``A simple optical demonstration of geometric phases
from multilayer stacks: the Wigner angle as an
anholonomy,''
J. Mod. Opt. \textbf{48}, 21-34 (2001).

\bibitem{geomphas}
A. Shapere and  F. Wilczek (Editors),
\textit{Geometric Phases in Physics}
(World Scientific, Singapore, 1989)

\bibitem{Ara97}
P. K. Aravind,
``The Wigner angle as an anholonomy in rapidity space,"
Am. J. Phys. \textbf{65},  634-636 (1997).

\end{thebibliography}
\end{document}